\documentclass[fleqn,usenatbib]{mnras}

\usepackage{newtxtext,newtxmath}

\usepackage[T1]{fontenc}

\DeclareRobustCommand{\VAN}[3]{#2}
\let\VANthebibliography\thebibliography
\def\thebibliography{\DeclareRobustCommand{\VAN}[3]{##3}\VANthebibliography}

\usepackage{graphicx}	
\usepackage{amsmath}	

\usepackage{psfrag,color}
\usepackage{color,soul} 
\usepackage{booktabs}
\usepackage{threeparttable}
\usepackage{graphicx}	
\usepackage{amsmath}	

\usepackage[utf8]{inputenc}

\newcommand{\Ne}{{$n_{\rm e}$}}
\newcommand{\Te}{{$T_{\rm e}$}}

\newcommand\W{{$\lambda$}}
\def\i{\,{\sc i}}

\newcommand{\oiii}{O\thinspace{\sc iii}} 
 
\newcommand{\nii}{N\thinspace{\sc ii}} 
\newcommand{\niv}{N\thinspace{\sc iv}} 
\newcommand{\niii}{N\thinspace{\sc iii}}

\newcommand{\siliii}{Si\thinspace{\sc iii}}
 
\newcommand{\oii}{O\thinspace{\sc ii}}

\newcommand{\ciii}{C\thinspace{\sc iii}}

\newcommand{\sii}{S\thinspace{\sc ii}}

\newcommand{\ariv}{Ar\thinspace{\sc iv}} 
\newcommand{\hi}{H\,{\sc i}}

\usepackage{orcidlink}

\newcommand{\srf}[1]{{\color{orange}Sophia: #1}}

\usepackage{threeparttable}

\title[High-z Galaxies: Insights from the \Te-Method]{A Self-Consistent Direct Method for Chemical Abundances in High-$z$ Galaxies with JWST
}

\author[K. Z. Arellano-C\'ordova]{
K. Z. Arellano-C\'ordova \orcidlink{0000-0002-2644-3518} $^{1}$\thanks{E-mail: ziboney@gmail.com and k.arellano@ed.ac.uk (KZAC)},
J. E. M\'endez-Delgado \orcidlink{0000-0002-6972-6411}$^{2}$\thanks{E-mail: jmendez@astro.unam.mx (JEMD)},
S. R. Flury\orcidlink{0000-0002-0159-2613}$^{1}$,\and
C. Esteban\orcidlink{0000-0002-5247-5943}$^{3,4}$,
K. Kreckel\orcidlink{0000-0001-6551-3091}$^{5}$,
J. Garc\'ia-Rojas,\orcidlink{0000-0002-6138-1869}$^{3,4}$,
F. Cullen\orcidlink{0000-0002-3736-476X}$^{1}$,
L. Carigi\orcidlink{0000-0002-2023-466X}$^{2}$,\and 
C. Morisset\orcidlink{0000-0001-5801-6724}$^{6,7}$, 
F. F. Rosales-Ortega\orcidlink{0000-0002-3642-9146}$^{8}$, 
A. Peimbert\orcidlink{0000-0001-7042-2207}$^{2}$,
T. M. Stanton\orcidlink{0000-0002-0827-9769}$^{1}$,
D. Scholte\orcidlink{0000-0002-6867-1244}$^{1}$\\
$^{1}$ Institute for Astronomy, University of Edinburgh, Royal Observatory, Edinburgh EH9 3HJ, UK \\
$^{2}$ Instituto de Astronom\'ia, Universidad Nacional Aut\'onoma de M\'exico, Apartado Postal 70-264, Coyoac\'an, 04510, Mexico City, Mexico.\\
$^{3}$ Instituto de Astrofísica de Canarias, 38205 La Laguna, Tenerife, Spain \\
$^{4}$ Departamento de Astrofísica, Universidad de La Laguna, 38206 La Laguna,Tenerife,Spain \\
$^{5}$ Astronomisches Rechen-Institut, Zentrum für Astronomie der Universität Heidelberg, Mönchhofstraße 12-14, D-69120 Heidelberg\\
$^{6}$ Universidad Nacional Aut\'onoma de M\'exico, Instituto de Astronom\'ia, AP 106, 22800 Ensenada, BC, M\'exico\\ 
$^{7}$ Universidad Nacional Aut\'onoma de M\'exico, Instituto de Ciencias F\'sicas, Av. Universidad s/n, 62210 Cuernavaca, Mor., M\'exico\\
$^{8}$ Instituto Nacional de Astrof\'isica, \'Optica y Electr\'onica (INAOE), Luis E. Erro 1, Tonantzintla, 72840 Puebla, Mexico
}

\date{Accepted XXX. Received YYY; in original form ZZZ}

\pubyear{2020}

\begin{document}
\label{firstpage}
\pagerange{\pageref{firstpage}--\pageref{lastpage}}
\maketitle

\begin{abstract}
The unprecedented rest-frame UV and optical coverage provided by JWST enables simultaneous constraints on the electron density (\Ne) and temperature (\Te) of ionized gas in galaxies at $z > 5$. We present a self-consistent direct method based on multiple \oiii] (\W1661,66) and [\oiii] (\W4363, and \W5007) transitions to characterize the physical conditions of the high-ionization zone. This new approach is insensitive to a wide range of \Ne\ due to the high critical densities of the \oiii] and [\oiii] transitions. Applying this technique to six galaxies at $z \sim 5$–9, we find electron densities up to \Ne\ $\sim 3 \times 10^{5}$ cm$^{-3}$ and temperatures of \Te\ $\sim 20,000$ K in systems at $z > 6$. Accounting for these self-consistent densities changes the derived \Te\ and modifies the inferred metallicities by up to 0.29~dex relative to previous estimates. We discuss the reported N/O overabundances in the high-$z$ galaxies from our sample, which arise entirely from the high N$^{3+}$/H$^{+}$ values inferred from \niv] lines. We point out that a \Te-stratification, in which the N$^{3+}$ zone has a slightly higher \Te\ than \Te([\oiii]), could substantially reduce the inferred N/O. Quantitatively, if \Te(N$^{3+}$) were 10\% higher than \Te([\oiii]), this could induce a systematic overestimation of N$^{3+}$/O$^{2+}$ of nearly 50\%.
Classical N/O diagnostics such as N$^{+}$/O$^{+}$, due to their critical densities, can significantly impact the inferred N/O abundance in the presence of high-density gas, whereas N$^{2+}$/O$^{2+}$ place these galaxies closer to $z\sim0$ systems in the N/O–O/H plane.
While promising, this method explicitly accounts for density inhomogeneities but does not incorporate the mathematical formalism of possible temperature fluctuations, which may also bias abundance determinations. Its accuracy ultimately depends on the reliability of UV atomic data at the high-density regime probed. Future JWST programs with larger and more diverse samples will be essential to test the universality and robustness of these results.
\end{abstract}



\section{Introduction} 
\label{sec:intro}

In recent years, the JWST has transformed our understanding of the distant Universe by enabling observations of cosmologically remote systems. Its unprecedented spectroscopic capabilities have allowed the determination of the physical conditions and chemical abundances of the youngest galactic systems, providing a unique opportunity to test long-standing theories on the origin and evolution of the cosmos. Instruments such as NIRSpec \citep{Jakobsen:22} have enabled the detection of multiple spectral line diagnostics across both the optical and ultraviolet wavelength ranges in galaxies at $z > 6$. This multiwavelength access has opened an unparalleled window for studying key physical properties, including the abundances of chemical elements such as O, He, Ne, N, S, and Ar \citep[e.g.,][]{arellanocordova22b,arellanocordova25,rogers24,Stanton:25,Berg:25, Rogers25} in the ionized gas of their star-forming regions.

The relative abundances of these elements allow us to probe the nucleosynthetic processes associated with stellar birth, life, and death. In particular, ratios such as N/O and C/O serve as cosmic clocks for constraining star-formation histories \citep{Matteucci:86,Woosley:02, Chiappini:03, Carigi:05}. Oxygen production is primarily linked to the deaths of massive stars ($M_{\star} \gtrsim 8 M_{\odot}$) as core-collapse supernovae (CCSNe), which enrich the interstellar medium (ISM) on relatively short timescales. By contrast, the nucleosynthetic production of  carbon and nitrogen also includes secondary contributions from low- and  intermediate-mass stars ($1 M_{\odot} \lesssim M_{\star} \lesssim 3 M_{\odot}$ and $3 M_{\odot} \lesssim M_{\star} \lesssim 8 M_{\odot}$, respectively), which are released into the ISM on significantly longer timescales \citep{kobayashi20, Romano2022}. Moreover, massive stars in binary interaction could contribute to the chemical enrichment \citep[e.g.,][]{famer23}. Therefore, diagnostic diagrams of N/O versus O/H and C/O versus O/H provide key insights into the evolutionary state of galaxies in the context of their star formation histories and initial mass functions \citep[e.g.][]{kobayashi24, Curti:25, arellanocordova25}.

Surprisingly, several studies of high-$z$ galaxies observed by the JWST have reported anomalously high N/O abundances (and, in some cases, C/O) for the low O/H abundances measured in these systems \citep{Cameron23, isobe23b,topping24a,Senchyna:24,arellanocordova25, morel25}. This putative nitrogen excess has led to the proposal of a variety of physical scenarios to explain these chemical anomalies, including stellar winds from very massive stars  \citep[VMS, $M_{\star} \gtrsim 100M_{\odot}$;][]{Vink:24,Nandal2024a}, a top-heavy initial mass function  \citep[IMF; e.g.,][]{curti25a}, the presence of super-massive stars \citep[SMS, $M_{\star} \gtrsim 1000M_{\odot}$;][]{marqueschaves:24}, Population III stars \citep{Nandal2024b}, stripped stars in globular clusters \citep{isobe23b}, stochastic star-formation histories characterized by multiple bursts interspersed with quiescent periods \citep[e.g.,][]{kobayashi24, Rossi24}, and the growth of early super-massive black holes \citep{maiolino24, Watanabe24}, among other scenarios that are relatively exotic when compared to conditions in the local universe \citep[e.g.,][]{Senchyna:24, Curti:25, morel25}.

Almost all high-$z$ galaxies with reported N/O versus O/H anomalies in the literature have been analyzed using the so-called ``direct method''. This method is widely regarded by many authors as the gold standard for chemical abundance determinations \citep[e.g][]{PerezMontero:17, peimbert17, kewley+19}. Although its application is highly inhomogeneous in practice, its fundamental requirement is the detection of at least one auroral or transauroral line, such as [\oiii]~$\lambda 4363$, which can be paired with the nebular line, such as [\oiii]~$\lambda 5007$, to have a \Te\ sensitive ratio.
When applied to high-$z$ galaxies, the same procedures developed for studies of the local universe are generally followed: first, the electron density ($n_{\rm e}$) is determined using intensity ratios of lines arising from atomic levels with similar energies but different collisional de-excitation rates, such as [\sii]~$\lambda 6716/\lambda 6731$ or \ciii]~$\lambda 1907/\lambda 1909$; next, \Te\ is estimated using line intensity ratios of transitions between atomic levels with different excitation energies, such as [\oiii]~$\lambda 4363/\lambda 5007$. Finally, values of \Ne\ and \Te\ are adopted to compute the ionic abundances of the species present in the emission spectrum.

Although it is called direct whenever a \Te-sensitive line is detected, this method often requires several approximations that are not so direct, such as combining diagnostics from different ions to simultaneously estimate \Ne\ and \Te, assuming ionic coexistence and homogeneous structure of densities and temperatures. 
For a line intensity ratio such as [\oiii]~$\lambda 4363/ \lambda 5007$ to serve as a pure temperature diagnostic, the electron density of the gas in the region where O$^{2+}$ coexists must be well below the density at which collisional de-excitation of [\oiii]~$\lambda 5007$ becomes significant. In practice, collisional suppression begins to affect the ratio at \Ne\ $\gtrsim$ $10^{5}$ cm$^{-3}$ and becomes important for \Ne\ of order of few $\times 10^5$ cm$^{-3}$.
While this condition is often satisfied in the local universe, the electron density in many high-$z$ galaxies could be considerably higher than this threshold \citep[e.g.,][]{Davies2021,isobe23a, Yanagisawa24b}. As a result, the adoption of a particular density value can significantly influence the \Te\ inferred from ratios such as [\oiii]~$\lambda 4363/ \lambda 5007$ \citep[e.g.,][]{marconi24, hayes25, martinez25}. 

This issue is further complicated by the fact that, since the pioneering studies of ionized gas, the existence of density variations has been well established \citep[e.g.,][]{osterbrock89}. Density and temperature variations occur in star-forming regions on spatial scales as small as milliparsecs, as revealed by observational studies with the Hubble Space Telescope (HST) \citep{ODell:2003,ODell:2017}, and likely on scales even smaller than the current resolution limit. These variations have also been detected on scales as large as a few tenths of parsecs \citep{MesaDelgado:2011, McLeod:2016, Royer:2025}. They introduce biases among the different diagnostics available in emission-line spectra \citep{Rubin:89}, which explains why most objects exhibit large discrepancies between the \Ne\ values obtained from different diagnostics \citep{Peimbert:71, MendezDelgado:23b, MendezDelgado:24a}. These density variations, which are clearly present in the local universe, are also likely to exist at high redshift, thereby introducing significant biases. Moreover, local low metallicity analog galaxies such as I~Zw~18 exhibit electron density gradients across the ionization structure of nebular gas \citep{Campbell1990}. The presence of density gradients further complicates application of the direct method, highlighting the necessity of measuring temperatures and densities from the same ionic species.

Fortunately, the wavelength coverage and sensitivity of JWST for galaxies at $z > 5$ allows us to detect not only [\oiii]~$\lambda \lambda 4363, 5007$ but also the \oiii]~$\lambda \lambda 1661, 1666$ doublet (hereafter \oiii]~$\lambda \lambda 1664$+), both originating from the $^5S_2$ level, which lies 2.12 eV higher in energy than the $^1S_0$ level that gives rise to [\oiii]~$\lambda 4363$ (See Fig.~\ref{fig:O3_Grotrian}). For densities greater than $10^{5}$ cm$^{-3}$, the \oiii]~\W 1666/[\oiii]~\W4363 ratio is significantly less sensitive to density than the classical [\oiii]~$\lambda 4363/\lambda 5007$ diagnostic, while still maintaining a strong dependence on \Te. This behavior is expected, given that the critical density of the $^5S_2$ level is nearly five orders of magnitude higher than that of the $^1D_2$ level, which produces the nebular [\oiii]~$\lambda 5007$ line as shown in Table~\ref{tab:O3_levels}. This enables us to simultaneously and self-consistently determine \Te\ and \Ne\ in high-density galaxies using the \oiii]~$\lambda 1666$/[\oiii]~$\lambda 4363$ and [\oiii]~$\lambda 4363/\lambda 5007$ ratios, respectively. With this method, the physical conditions associated with the highly ionized gas, such as O$^{2+}$, can be determined in a truly direct way. Consequently, the abundance of O$^{2+}$ --which represents the largest ionic fraction under the high-ionization conditions typical of the early universe-- can be determined in a genuinely direct manner.

For this reason, in this work we compiled a sample of  high-$z$ galaxies with simultaneous detections of \oiii]~$\lambda \lambda 1664+$, and [\oiii]~$\lambda \lambda 4363, 5007$ observed with JWST. Three of these galaxies exhibit anomalously high N/O abundances. However, in this study we focus on those galaxies with the detections of \ion{O}{iii}] and [\ion{O}{iii}]. In this context, Our goal is to recalculate their physical and chemical conditions in a fully self-consistent manner using the method described in the previous paragraph.

This paper is organized as follows. In Sec.\ref{sec:sample}, we describe the sample of high-$z$ galaxies analyzed in this work, summarizing their physical conditions and chemical abundances as reported in the original studies. In Sec.\ref{sec:physica_conditions}, we present our self-consistent method to derive \Te\ and \Ne, along with the ionic and total abundances of oxygen and nitrogen. In Sec.\ref{sec:disc}, we discuss the metallicities and N/O ratios of the high-$z$ sample in comparison with local star-forming objects, and place these results in the context of key scaling relations such as the mass–metallicity relation and the N/O–O/H relation. Finally, in Sec.\ref{sec:conclusion}, we summarize our conclusions and provide some perspectives for future work.

\section{JWST Observational Sample}
\label{sec:sample}

We have reviewed the literature for high-$z$ galaxies with JWST/NIRSpec detection of the auroral [\oiii]$~\lambda$4363, \oiii]$~\lambda\lambda$1664+, and nebular [\oiii]~$\lambda$5007 lines. These lines are crucial because the combination of their relative fluxes allows us to derive \Ne\ and \Te\ associated with the high-ionization structure of the gas in these galaxies.

 (1) {\it EXCELS-121806} ($z = 5.23$), this target is part of the EXCELS program \citep{carnall24} observed with NIRSpec/JWST. \citet{arellanocordova25} derived a metallicity of 12 + log(O/H) = $7.97^{+0.05}_{-0.04}$ and \Te([\oiii]) = $14 900 \pm 1100$ K from the [\oiii]~$\lambda$4363/$\lambda$5007 ratio, and \Ne([\ion{S}{ii}]) = $600^{+920}_{-400}$ cm$^{-3}$. \citep[e.g.,][]{arellanocordova25} found a high nitrogen abundance of log(N/O) = $-0.86^{+0.15}_{-0.11}$, consistent with chemical evolution models in which N enrichment is produced by intermediate-mass stars (e.g., AGB stars).

(2) {\it RXCJ2248-ID} ($z = 6.2$), \citet{topping24a} showed the results of an extremely compact system ($<200$ pc) using NIRSpec/JWST observations. Its UV and optical spectra show high-ionization features such as \ion{He}{ii}, \ion{C}{iv}, and \ion{N}{iv}], consistent with an extreme ionization parameter ([\oiii]/[\oii] = 189) and a sub-solar metallicity of 12 + log(O/H) = $7.43^{+0.17}_{-0.09}$. The density structure, traced by UV \ion{Si}{iii}], [\ion{C}{iii}], and \ion{N}{iv}] diagnostics, indicates very high densities ($10^{4}$–$10^{5}$ cm$^{-3}$). These authors derived \Te([\oiii]) = $24600 \pm 2600$ K from [\oiii]$~\lambda$4363/$\lambda$5007 and \Te([\oiii]) = $23300 \pm 4,100$ K from \oiii]~$\lambda \lambda 1664^+$/[\oiii]~$\lambda 4363$ assuming a \Ne = 10$^{4}$ cm$^{-3}$. Using \niii]$~\lambda$1750 and \ion{N}{iv}]$~\lambda\lambda$1483,86, they measured log(N/O) = $-0.39^{+0.11}_{-0.10}$, suggesting rapid nitrogen enrichment during an intense star-formation phase. 

Recently, \citet{Berg25b} presented deeper, high-resolution optical observations of RXCJ2248-ID (and identified as RXCJ2248-ID3 by the authors), reporting a complex kinematic structure for this galaxy. From their chemical analysis, the authors obtained a similar value of $\langle\log(\mathrm{N/O})\rangle = -0.390 \pm 0.035$ than in \citet{topping24a}, but a higher value of $12 + \log(\mathrm{O/H}) = 7.749 \pm 0.023$. These authors adopted \Ne([\niv]) = $2.65^{+0.45}_{-0.38}$ $\times$10$^{5}$ cm$^{-3}$ as their preferred density to infer \Te([\oiii]) = $19,700 \pm 300$ K. Moreover, for the ionic abundances, \citet{Berg25b} used the various density diagnostics derived for RXCJ2248-ID3 according to the ionization potentials of O$^{+}$ and N$^{+}$, for which \Ne(\siliii]) was adopted. For O$^{2+}$, N$^{2+}$, and N$^{3+}$, \Ne(\ciii]) and \Ne([\ariv]) were preferred, with densities ranging around $10^{4}$ cm$^{-3}$, respectively.
For this study, we use the fluxes reported independently by \citet{topping24a} and \citet{Berg25b} for this galaxy, which we refer to as RXCJ2248-ID-T24 and RXCJ2248-ID3-B25, respectively.

(3) {\it A1703-zd6} ($z = 7.04$), a lensed galaxy analyzed by \citet{topping25a} with NIRSpec. High densities, $(8$–$19) \times 10^{4}$ cm$^{-3}$, were measured from [\ion{C}{iii}]$~\lambda\lambda$1906,09 and \ion{N}{iv}]$~\lambda\lambda$1483,86. A low metallicity of 12 + log(O/H) = $7.47 \pm 0.17$ was obtained from \Te([\oiii]) = $23000 \pm 3200$ K using the [\oiii]~$\lambda$4363/$\lambda$5007 diagnostic. Detections of \niii] and \ion{N}{iv}] yield a high log(N/O) = $0.6 \pm 0.3$, placing A1703-zd6 among the $z > 6$ galaxies with very high N/O abundances in the early Universe.

(4) {\it JADES-GS-z9-0} ($z = 9.43$), the ISM conditions of this compact galaxy were presented by \citet{curti25a}. The young stellar population and high ionization parameter revealed by NIRCam and NIRSpec observations indicate a metal-poor system with a metallicity of 12 + log(O/H) = $7.49 \pm 0.11$.
They report a metallicity of 12 + log(O/H) = $7.49 \pm 0.11$, which is quite low due to their high \Te\ = $20137 \pm 1940$ K inferred from the [\oiii]$~\lambda$4363/$\lambda$5007 diagnostic and comparable to that from \oiii]~$\lambda \lambda 1664^+$/[\oiii]~$\lambda 4363$. \citet{curti25a} also reported \Ne([\ion{S}{ii}]) = $670 \pm 430$ cm$^{-3}$. Based on the $3\sigma$ lower limit of the \niv] $\lambda$1483/1486 ratio, \citet{curti25a} concluded that this galaxy does not exhibit high-density gas. These authors also derived the N/O ratio using \niii]$~\lambda\lambda$1750 in combination with \ion{C}{iii}]~$\lambda$1909 (C$^{++}$/N$^{++}$), taking advantage of their similar range in ionization potentials ($\sim$ 20-48 eV). \citet{curti25a} reported log(N/O) = $-0.93 \pm 0.24$, and their chemical evolution modeling suggests that a single burst of star formation, combined with massive stars regulated by a top-heavy IMF, can explain the observed chemical abundance patterns.

(5) {\it CEERS-1019} ($z = 8.6782$), the physical conditions and chemical abundances of this galaxy were analyzed by \citet{marqueschaves:24}, who used a spectroscopic combination of NIRSpec/PRISM and NIRSpec/G395M to cover the \oiii]~$\lambda1664^+$ and [\oiii]~$\lambda\lambda4363, 5007$ lines, respectively \citep[see also][]{larson23, isobe23b}. It is important to note that the NIRSpec/PRISM and NIRSpec/G395M spectra were not renormalized using the flux of common emission lines; instead, a function was fitted to match the spectral continua. This combination may increase the systematic errors arising from flux calibration between the rest-frame UV and optical regimes. The authors reported an electron temperature of \Te~from the [\oiii]~$\lambda4363/\lambda5007$ ratio of $18850 \pm 3250$ K, which leads to an 12+log(O/H) = $7.70 \pm 0.18$. Likewise, these authors reported a rather elevated log(N/O) abundance, up to $-0.13$, primarily driven by the high N$^{3+}$/H$^{+}$ abundance they derived.

Overall, the sample of high-$z$ galaxies discussed here is reported to show very high \Ne, low metallicities, and high N/O ratios, with the exception of EXCELS-121806, whose properties appear more comparable to those of local galaxies \citep[e.g.,][]{arellanocordova25, arellanocordova25b}. The gas-phase properties of these systems are generally derived from [\oiii]~$\lambda$4363/$\lambda$5007 or \oiii]~$\lambda1664^+$/[\oiii]$\lambda$5007 ratios, yielding \Te([\oiii]) values $> 20,000$ K for $z > 5$ galaxies, under the density structures assumed in the original analyses.

We are aware that several studies report combinations of genuine detections of some of the [\oiii] lines of interest together with upper limits for others, or require the drastic mixing of different instruments and/or gratings in different epochs to include both the rest-frame UV and optical [\oiii] detections \citep{Bunker:23, pascale23, schaerer24, alvarez-marquez25, naidu25}. We consider that the systematic uncertainties associated with objects observed under such conditions are too large to be reliably analyzed with the methodology proposed in this work. The exception is the case of CEERS-1019, that combines NIRSpec/PRIS and NIRSPEC/G395M spectroscopy which we include precisely to test the potential observational uncertainties we may encounter. The results from these studies will be discussed in Sec.~\ref{sec:disc}.

\section{Physical Conditions and Chemical Abundances}
\label{sec:physica_conditions}

The determination of extinction in high-$z$ galaxies represents a major challenge, particularly at UV wavelengths, where extinction curves in the local universe often show an abrupt feature or ``bump'' \citep{Costero:1970,Cardelli:89,ODonnell:94}. This bump is associated with the presence of polycyclic aromatic hydrocarbons (PAHs) and other chemical compounds whose prevalence and abundance in high-$z$ galaxies remain uncertain \citep{Draine:2003,Reddy:2025}. At optical wavelengths, where extinction curves tend to be smoother and more consistent with one another, extinction values derived from the Balmer decrement in high-$z$ galaxies are generally found to be very small. In fact, in some cases the H$\alpha$/H$\beta$ ratios yield values slightly lower than the theoretical Case-B prediction. This may indicate significant leakage of ionizing photons from these galaxies, potentially contributing to the reionization of the early universe \citep[e.g.,][]{Scarlata:2024, flury22a, Yanagisawa24, McClymont25, arellanocordova25}.

In our analysis, we adopt observed fluxes without extinction correction, under the assumption that extinction is minimal, consistent with Balmer decrement measurements and the methodology commonly followed in the literature. However, this does not imply that extinction has no significant impact on the physical and chemical conditions of high-$z$ galaxies. Our goal here is simply to test if a careful treatment of electron density and temperature alone can already induce substantial changes in the inferred chemical abundances of the gas, thereby accounting for some of the reported chemical anomalies. This, of course, does not exclude changes in the typical extinction treatment as an important factor to consider, but rather highlights that it introduces additional, poorly constrained degrees of freedom. This strengthens the central point of our analysis: empirical abundance determinations in high-$z$ galaxies may not be sufficiently robust, and variations in the underlying physical conditions (including extinction) may offer more plausible explanations for the reported chemical anomalies than exotic nucleosynthetic enrichment scenarios.

\begin{figure*}
    \centering
    \includegraphics[width=0.95\linewidth]{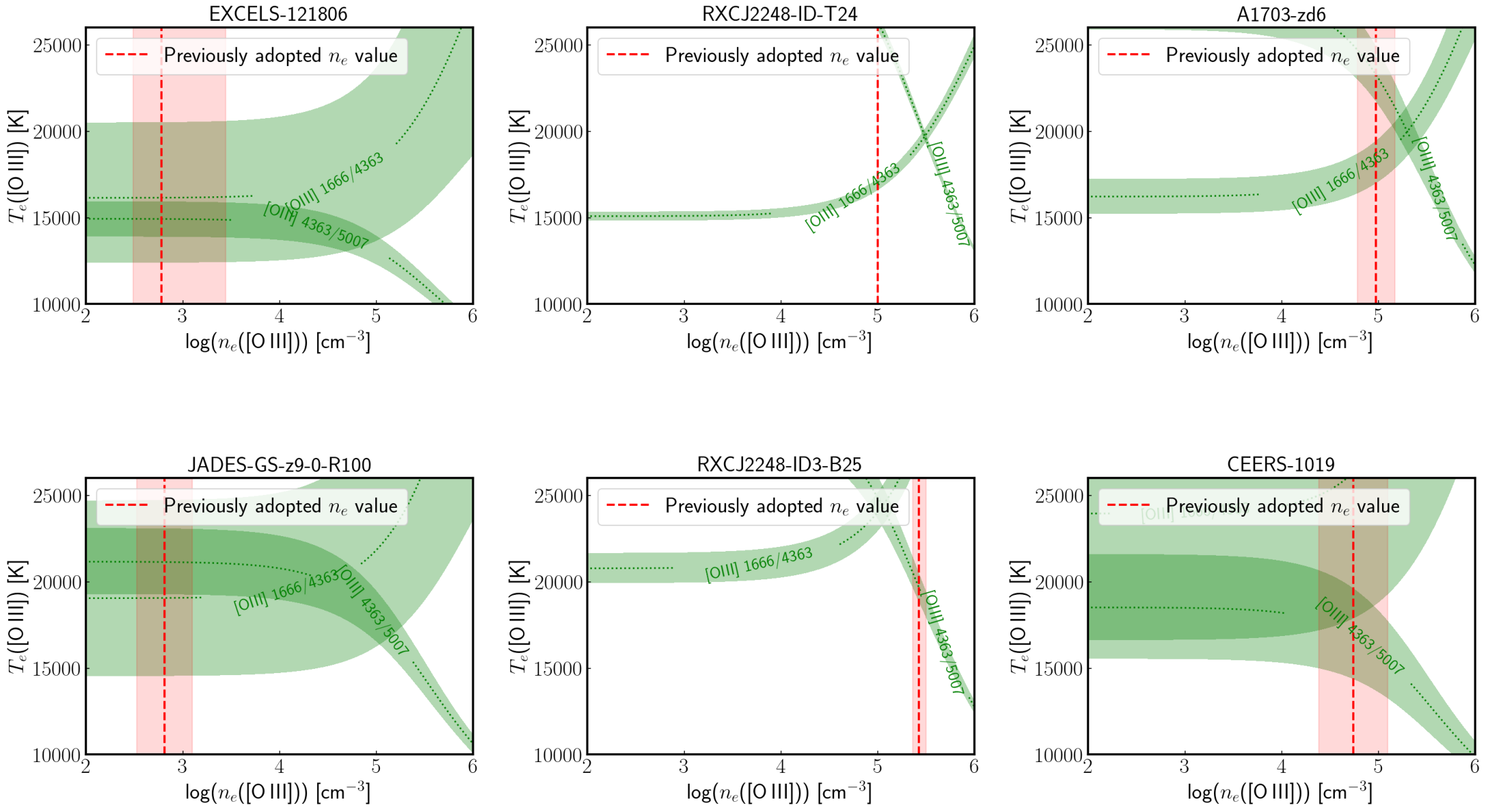}
    \caption{ The \Te\ \textit{vs.} \Ne\ diagnostics based on the reported \oiii]~$\lambda 1666$ and [\oiii]~$\lambda \lambda 4363, 5007$ fluxes and their associated uncertainties for the sample of high-$z$ galaxies listed in Table~\ref{tab:oiii_properties}. The most likely combination of \Ne\ and \Te\ for the high-ionization gas is indicated by the intersection of the diagnostic curves. When the errors are small, the combination of these diagnostics provide a self-consistent approach to derive \Te([\oiii]) and \Ne([\oiii]). The density values previously adopted to determine \Te([\oiii]) are indicated by the red vertical line, with a shaded band representing the uncertainty estimated by the reference authors. In the case of RXCJ2248-ID-T24, \citet{topping24a} only stated that a density of $1\times 10^{5}$ cm$^{-3}$ was assumed, without providing a clear associated uncertainty range.}
    \label{fig:plasmaplots}
\end{figure*}

We calculate the physical and chemical conditions of the sample using the fluxes reported by the reference authors with {\tt PyNeb} \citep[version 1.1.14][]{luridiana15,Morisset:20}, together with the selected atomic data shown in Table~\ref{tab:atomic_data} for N and O ions. To illustrate the dependence of \oiii]~$\lambda 1666$/[\oiii]~$\lambda4363$ and [\oiii]~$\lambda 4363/\lambda5007$ on \Ne\ and \Te, we show  the \Te-\Ne diagnostic diagrams for each galaxy in Figure~\ref{fig:plasmaplots} generated using the {\tt PyNeb} routine {\tt getEmisGridDict}. 
The intersection of the curves represents the most likely combination of \Ne\ and \Te\ for the ionized volume traced by the O$^{2+}$ ion. The resulting \Ne\ and \Te values are reported in Table~\ref{tab:oiii_properties}. Additionally, we run a statistical test using a Monte Carlo simulation which samples the distribution of possible temperatures and densities given the flux ratios of \oiii]~\W 1666/[\oiii]\W4363 and [\oiii]~$\lambda 4363/\lambda5007$.  As a check on our methodology, we forward-model the observations by MCMC sampling \Te\ and \Ne\ such that these minimize the difference between the {\tt PyNeb} emissivity ratios and the measured line ratios (see Appendix \ref{sec:MC_simulations}). We present these results in Fig.~\ref{fig:MC_simulations}, which are in agreement with the \Te-\Ne\ diagnostics of Fig.~\ref{fig:plasmaplots}. 

\begin{table*}
\centering
\caption{Electron temperature and density, metallicities and N/O abundance ratios.} 
\label{tab:oiii_properties}
\begin{tabular}{l c c c c c c}
\toprule
Property & EXCELS-121806  & RXCJ2248-ID-T24 & RXCJ2248-ID3-B25$\dagger$ & A1703-zd6 & JADES-GS-z9-0 & CEERS-1019\\
\midrule
$z$ & 5.23  & 6.1 & 6.1 & 7.04 & 9.43 & 8.68 \\
log(M$_{\star}$/M$_{\odot}$) & $8.02\pm0.06$ & $8.05\pm0.17$ & $ 7.66^{+1.99}  _{-0.27} $ & $7.70\pm0.24$  & $8.18\pm0.06$ & $9.30\pm 0.13$\\
\Te([\oiii]) [K] & $14\,900 \pm 1\,650$ & $19\,700 \pm 500$  & $24\,280 \pm 1000$ & $19\,900 \pm 1\,300$ & $20\,000 \pm 800$ & $18\,340 \pm 3200$\\
\Ne([\sii]) [cm$^{-3}$] &$600^{+920}_{-400}$ & --  & -- & --  & $650\pm430$ &--\\
\Ne([\oiii]) [$\times 10^{5}$ cm$^{-3}$] & $\leq 1$ & $3.02 \pm 0.07$ & $1.11 \pm 0.3$ &$2.00^{+0.33}_{-0.54}$ & $\leq 0.4$& $\leq 1$\\
O$^{+}$/H$^{+}$ [$\times 10^{5}$] & $1.35 \pm 0.70$ & $0.07 \pm 0.03$ & $0.04 \pm 0.03$ & --- & $0.09 \pm 0.05$ & $1.50 \pm 0.88$\\
O$^{2+}$/H$^{+}$ [$\times 10^{5}$] & $8.44 \pm 2.95$ & $5.22 \pm 0.29$ & $3.01 \pm 0.35$ & $4.58 \pm 0.75$ & $2.52 \pm 0.29$ & $5.46 \pm 2.67$ \\
12+log(O/H) & $7.99 \pm 0.13$ & $7.72 \pm 0.03$  & $7.50 \pm 0.05$ & $7.66 \pm 0.07$ & $7.42 \pm 0.05$ & $7.74 \pm 0.22$ \\
$\Delta$(O/H)$\dagger$ & $0.02$ &  $0.29$ & $-0.25$ & $0.19$ & $0.02$ &  0.04\\
N$^{+}$/H$^{+}$ [$\times 10^{5}$] & $0.15\pm0.04$ & $0.03\pm0.01$ & $0.028\pm0.004$ & --- & ---& --- \\
N$^{2+}$/H$^{+}$ [$\times 10^{5}$] & --- & $0.30\pm0.04$ & $0.45\pm0.09$ & $0.36\pm0.18$ & $0.27\pm0.11$ & $0.80\pm1.69$\\
N$^{3+}$/H$^{+}$ [$\times 10^{5}$] & --- & $1.55\pm0.16$ &$1.16\pm0.18$ & $0.74\pm0.26$ & ---& $3.15\pm1.66$ \\
log(N$^{+}$/O$^{+}$) & $-0.96 \pm 0.25$ & $-0.30 \pm 0.21$ & $-0.18\pm0.26$& --- & ---& ---\\
log(N$^{2+}$/O$^{2+}$) & ---  & $-1.25 \pm 0.08$ & $-0.85 \pm 0.12$& $-1.13^{+0.28}_{-0.17}$ & $-0.97 \pm 0.21$ & $-0.88 \pm 0.34$\\
log((N$^{2+}$+ N$^{3+}$)/O$^{2+}$) & ---& $-0.46\pm0.07$ & $-0.30\pm0.11$ & $-0.65^{+0.22}_{-0.13}$ & --- & $-0.18\pm0.24$\\
log(N/O)$\ddagger$ & --- & $-0.45\pm0.05$ & $-0.28\pm0.07$& --- & ---& ---\\
\bottomrule
\end{tabular}
\begin{description}
    \item We have used \Ne([\ion{S}{ii}]) to determine O$^{+}$ and N$^{+}$. For RXCJ2248-ID-T24, RXCJ2248-ID3-B25, A1703-zd6, and CEERS-1019, we assumed \Ne$=10^{4}$ cm$^{-3}$ (see also Sec.~\ref{sec:physica_conditions}). Using \Ne([\oiii]), we obtained log(N$^{+}$/O$^{+}$) $\approx$ log(N/O) = $-1.00 \pm 0.05$ for RXCJ2248-ID-T24. 
     $^{\ddagger}$ Total N/O derived using log(N/O) = log((N$^{+}$+ N$^{2+}$ + N$^{3+}$)/(O$^{+}$ + O$^{2+}$)) \citep[see also][]{arellanocordova25}. $\dagger$ Differences in O/H between the metallicity derived in this study and the inferred from the original studies. $\dagger$These results use the reddening-corrected fluxes from the narrow component, adopting $E(B-V) = 0.053$ reported in \citet{Berg25b}, which suggests a low dust content. The redshifts and stellar masses are those reported in the original studies of each object. 
\end{description}
\end{table*}

Once \Ne\ and \Te\ are determined self-consistently using only \oiii] and [\oiii]  lines, we derive the ionic abundances using the {\tt getIonAbundance} routine from {\tt PyNeb}. For high-ionization species such as O$^{2+}$, N$^{2+}$, and N$^{3+}$, we directly adopt \Ne([\oiii]) and \Te([\oiii]) . For these determinations, we use the reported fluxes of [\oiii]$~\lambda 5007$, \niii]~$\lambda \lambda 1750$, and \niv]~\W\W1485. It is important to note that the choice of the [\oiii] line used to derive the ionic abundance of O$^{2+}$ is not critical, since the physical conditions are determined self-consistently and all lines yield the same results. For low-ionization species such as O$^{+}$ and N$^{+}$, we use the temperature relations determined by \citet{garnett92} under local-universe conditions and adopt the electron density reported by the reference authors using the classical [\sii]~$\lambda 6731/6717$ diagnostic for EXCELS-121806 and JADES-GS-z9-0. For RXCJ2248-ID-T24, RXCJ2248-ID3-B25, and CEERS-1019, we have adopted \Ne$=1\times10^{4}$~cm$^{-3}$, based mainly on the results of other diagnostics from \siliii] and \ciii] tracing high density gas \citep[see e.g.,][]{topping24a, Berg25b}.
For these ionic abundances we adopted the reported fluxes of [\oii]~$\lambda \lambda 3726,29$ and [\nii]~$\lambda 6584$. The adopted physical conditions for the low ionization ions are discussed in detail in Sec.~\ref{sec:disc}. 

To derive the total oxygen abundance (O/H), we added the contributions of O$^{+}$/H$^{+}$ and O$^{2+}$/H$^{+}$. Note in Table~\ref{tab:oiii_properties} that O$^{+}$ has a minimal impact on the total O/H abundance. For N/O, we derived the abundance ratio in four different ways: (1) for galaxies with the optical [\nii]\W6584 line, we use N$^{+}$/O$^{+}$ $\approx$ N/O based on the similar ionization potentials of these ions \citep{peimbert69}; (2) we also considered N$^{2+}$/O$^{2+}$ $\approx$ N/O using \niii]\W1750 and \oiii]\W1666; and (3) we included the contribution of \ion{N}{iv}]\W\W1485 as (N$^{2+}$ + N$^{3+}$)/O$^{2+}$ $\approx$ N/O. It is important to note that the latter abundance ratio may be overestimated, since N$^{3+}$ has a higher ionization potential (47.45–77.47 eV) than O$^{2+}$ (35.12–54.94). 
For RXCJ2248-ID we derived the total N/O abundance as N/O = (N$^{+}$ + N$^{2+}$ + N$^{3+}$)/(O$^{+}$ + O$^{2+}$). In Table~\ref{tab:oiii_properties}, we list the physical conditions as well as the ionic and total abundances of O and N for the high-$z$ sample.

\section{Results and discussion}
\label{sec:disc}

In this study, we derive \Ne\ and \Te\ for four galaxies at $z > 5$ using exclusively the different transitions of the O$^{2+}$ ion in both the rest-frame optical and UV (see Fig.~\ref{fig:O3_Grotrian}).
In Fig.\ref{fig:plasmaplots}, we show the utility of combining the \oiii]$~\lambda 1666$/[\oiii]$~\lambda 4363$ and [\oiii]~$\lambda 4363/\lambda 5007$ diagnostics to simultaneously and self-consistently determine \Ne\ and \Te\ in high-$z$ galaxies observed with JWST, respectively. This work presents the first determinations of \Ne([\oiii]) for galaxies at $z>5$. Overall, our methodology yields high electron densities on the order of $10^{5}$ cm$^{-3}$, significantly higher than the values adopted in some of the reference studies, with the exception of RXCJ2248-ID3 by \citet{Berg25b}, where they overestimate the density. As a result, the temperatures inferred from classical ratios such as [\oiii]$\lambda 4363/\lambda 5007$ can substantially differ from the previously reported values, leading to correspondingly different heavy-element abundances derived from collisionally excited lines (CELs).

We derive \Ne([\oiii]) values in the range $1.88$–$3.02 \times 10^{5}$ cm$^{-3}$ for RXCJ2248-ID-T24, A1703-zd6, and RXCJ2248-ID3-B25 ($z > 6$ galaxies; see Table~\ref{tab:oiii_properties}). For RXCJ2248-ID-T24, we obtain \Ne([\oiii]) = $3.02 \times 10^{5}$ cm$^{-3}$, consistent with the \Ne(\ion{N}{iv}]) reported by \citet{topping24a}. For RXCJ2248-ID3-B25, we find a value of  \Ne([\oiii]) = $1.11 \times 10^{5}$ cm$^{-3}$,  lower than that determined from \Ne(\ion{N}{iv}]) by \citet{Berg25b}. For A1703-zd6, we find a similarly high value, \Ne([\oiii]) = $2.00 \times 10^{5}$ cm$^{-3}$. For galaxies JADES-GS-z9-0, EXCELS-121806, and CEERS-1019, only upper limits to the density could be established, likely indicating that the uncertainties in the line measurements, particularly in \oiii]~$\lambda\lambda1664^+$, are high enough to prevent us from fully applying our method. In the case of CEERS-1019, the \oiii]~$\lambda\lambda1664^+$ line appears to be slightly overestimated since the colored bands are not fully contained within one another as in the other cases, probably due to the combination of UV data from NIRSpec/PRISM with those optical from NIRSpec/G395M. Although the measurement remains within the formal uncertainty range, if the rest-frame UV values are systematically overestimated with respect to the optical ones, this could introduce distortions in the derived physical conditions or chemical abundances that rely on the combination of UV-optical wavelengths. The case of these three galaxies highlights the need for deep observations with careful calibration across the full wavelength range in order to apply the self-consistent methodology presented in this work. Interestingly, for JADES-GS-z9-0 there is a tentative detection of \ion{N}{iv}]~$\lambda$1486, that in pair with \ion{N}{iv}]~$\lambda$1483/\ion{N}{iv}]~$\lambda$1486 suggest no significant contribution of extremely high-density regions ($>10^4$ cm$^{-3}$) \citep{curti25a}. However, no other UV density diagnostics are available for this galaxy. 

Based on our \Ne([\oiii]) determinations, we find \Te([\oiii]) $\sim 20,000$ K for four of the spectra at $z > 6$, while EXCELS-121806  yield \Te([\oiii]) = $14,900\pm1650$ K. For the galaxies at $z>6$, we find \Te([\oiii]) values up to $\sim 4,000$ K lower than those reported in the original studies with the exception of RXCJ2248-ID3-B25, where the opposite is the case, probably due to the an overestimate of \Ne\ used to infer \Te. These different temperatures have important implications for the ionic abundances of O$^{2+}$ and metallicity as well as other high-ionization ions. Indirectly, the determinations of N$^{+}$ and O$^{+}$ are also affected, since temperature relations are used to estimate the low-ionization \Te. However, when the O$^{+}$/O$^{2+}$ ratio is small, the impact of this effect is minimal, as shown in Table~\ref{tab:oiii_properties}.

\subsection{Density biases in the mass-metallicity relation}

\begin{figure}
    \centering
    \includegraphics[width=0.98\linewidth]{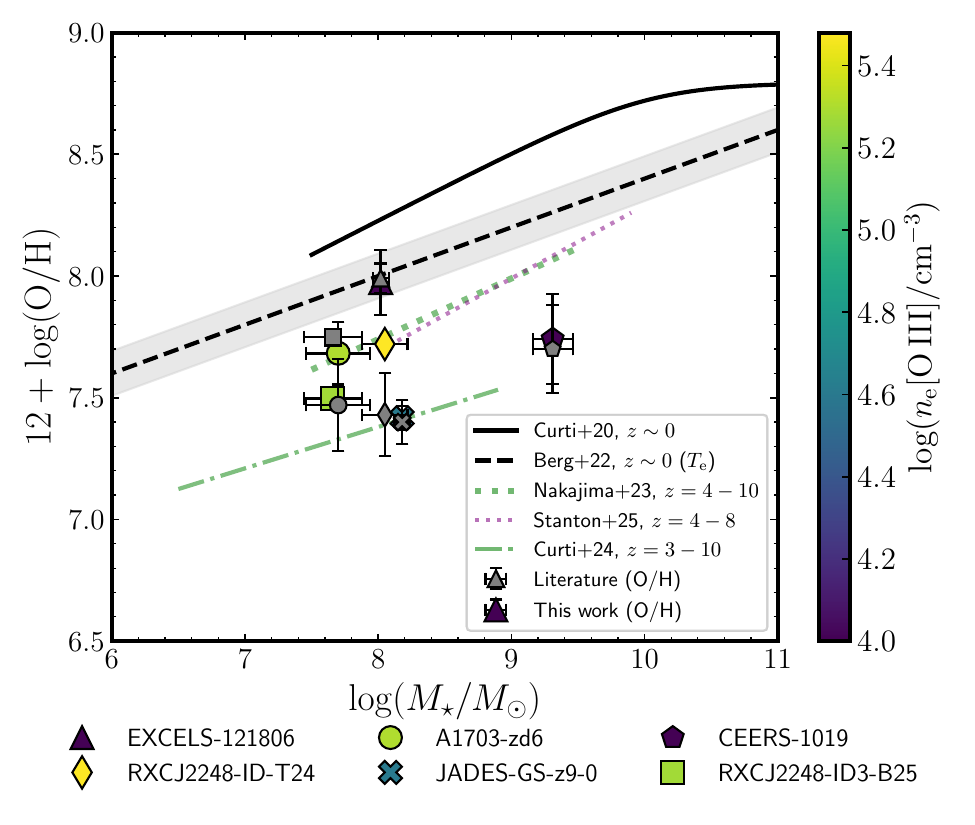}
    \caption{The mass–metallicity relation (MZR) for galaxies at $z=5$–$9$. Metallicity estimates derived in this study using the self-consistent method for \Te([\oiii]) and \Ne([\oiii]) are plotted in different symbols in colour-coded with \Ne, while those compiled from the original studies are shown with the same symbols in gray \citep{ topping24a, topping25a, arellanocordova25, curti25a, marqueschaves:24, Berg25b}. For RXCJ2248-ID-T24 and A1703-zd6 the metallicities obtained here are 0.18–0.29 dex higher than the values reported in the original works (see also Table~\ref{tab:oiii_properties}). For RXCJ2248-ID3-B25, we inferred a O/H 0.25 lower that in the original result. For comparison, different MZR determinations across various redshifts with metallicities estimated using strong-line methods are shown as lines : \citet[][$z\sim0$]{curti20}, \citet[][$z=3$–$10$]{curti24a}, and \citet[][$z=4$–$10$]{nakajima23}, \citet{stanton25b} ($z = 4-8$), and \citet[][$z\sim0$]{Berg22} using the \Te-sensitive method.}
    \label{fig:MZR}
\end{figure}

The systematic overestimation of \Te\ caused by underestimating \Ne\ leads to a corresponding underestimation of heavy-element abundances derived from CELs and viceversa. If the impact of \Ne\ variations increases at higher redshift, this could produce apparent variations in the mass–metallicity relation (MZR) that go beyond genuine cosmological differences.
In Table~\ref{tab:oiii_properties}, we present the metallicities obtained for the four high-$z$ galaxies studied here. Overall, the sample lies within the low-metallicity range, $7.42 < 12 + \log(\mathrm{O/H}) < 7.99$. Our results imply a shift of $\sim 0.19-0.29$ dex for RXCJ2248-ID-T24 and A1703-zd6 with respect to the previously reported values, whereas for EXCELS-121806 and JADES-GS-z9-0 the metallicity differences are minimal, on the order of $\sim 0.02$ dex, while that for CEERS-1019 we report a difference of 0.04 dex relative to the previous value. In the case of RXCJ2248-ID-B25, the metallicity shifts in the opposite direction, becoming  0.25 dex lower than that reported by \citet{Berg25b}. They adopted the density from \Ne([\niv]), which is higher than \Ne([\oiii]), leading to an underestimate of the electron temperature in the reference study.

Other studies have reported an underestimation of metallicities when incorrect or constant density values are assumed. For example, \citet{hayes25} reported that assuming \Ne\ $\sim 10^{6}$ cm$^{-3}$ decreases the value of \Te([\oiii]) in stacked spectra of high-$z$ galaxies due to the collisional de-excitation of [\oiii] $\lambda5007$, which has a direct impact on the derived metallicity. These authors found that the metallicity is 0.5 dex higher than when a low density value is assumed. Recently, \citet{martinez25} analyzed the gas conditions in individual spectra of local and high-$z$ galaxies assuming a multiphase density structure. These authors found that the metallicities of these galaxies can be underestimated by up to 0.67 dex when an inadequate gas density is used to derive \Te([\oiii]). 
The results of \citet{hayes25} and \citet{martinez25} rely on \Ne\ values derived from UV diagnostics (e.g., \ion{N}{iv}] or \ion{C}{iii}]) or on assumptions based on such diagnostics that could be not associated with the O$^{2+}$ emitting volume. 
Therefore, the methodology presented here allows us to alleviate the density dependence in the determination of \Te([\oiii]), providing robust measurements of this key parameter and metallicities in the presence of high-density gas.

In Fig.~\ref{fig:MZR}, we show the MZR for the analyzed sample. For comparison, we also plot the metallicities of the galaxies as reported in the original studies in gray and different fits to the MZR reported in the literature across redshifts \citep[e.g.,][]{curti20, Berg22, nakajima23, curti24a, stanton25b}. We find a significant offset toward higher metallicities, for the $z > 6$ galaxies (RXCJ2248-ID-T24, and A1703-zd6), which now fall closer to the parameter space defined by local systems. On the other hand, the results obtained with this methodology for EXCELS-121806 and JADES-GS-z9-0 are consistent with those derived from the traditional [\oiii]~$\lambda5007$/$\lambda4363$ diagnostic in the low-density limit. However, our approach is far more accurate for inferring \Te([\oiii]) at gas densities approaching the critical density of the lowest energy-state transitions of [\oiii]~$\lambda5007$. In other words, the methodology we propose is robust to the effects of gas density. As a result, our method not only serves as a powerful probe of both temperature and density simultaneously, but also as the only means to obtain accurate estimates of ionic abundances in extreme environments such as those at high redshift. These results in Fig.~\ref{fig:MZR} suggest that, although high-$z$ galaxies may indeed undergo genuine cosmological evolution in the MZR, their properties might not be as different from those of local objects as initially reported in the first JWST-based studies. However, a larger sample of galaxies with both \oiii] and [\oiii] detections is necessary to study the impact of high-density regions on shaping the MZR when \Te([\oiii]) and \Ne([\oiii]) are derived.

\subsection{Biases in the N/O-O/H relation}

One of the intriguing chemical properties of these galaxies is their claimed evidence of nitrogen enrichment. The N/O abundances reported for two of these systems are higher than those of local SFGs at similar metallicities \citep[e.g.,][]{topping24a, topping25a, marqueschaves:24, schaerer24}. In Table~\ref{tab:oiii_properties}, we also list the $\log(\mathrm{N/O})$ values derived in four different ways, depending on the available information for each galaxy. For a comparison between N/O and O/H in the local and high-$z$ universe, we adopt the DEep Spectra of Ionized REgions Database (DESIRED) sample of local star-forming regions \citep{MendezDelgado:23b, esteban25}. This represents the largest compilation of ionized regions with direct temperature determinations and a careful treatment of observational uncertainties in the spectra, allowing for highly precise chemical abundance measurements.

\begin{figure*}
    \centering
    \includegraphics[width=0.90\linewidth]{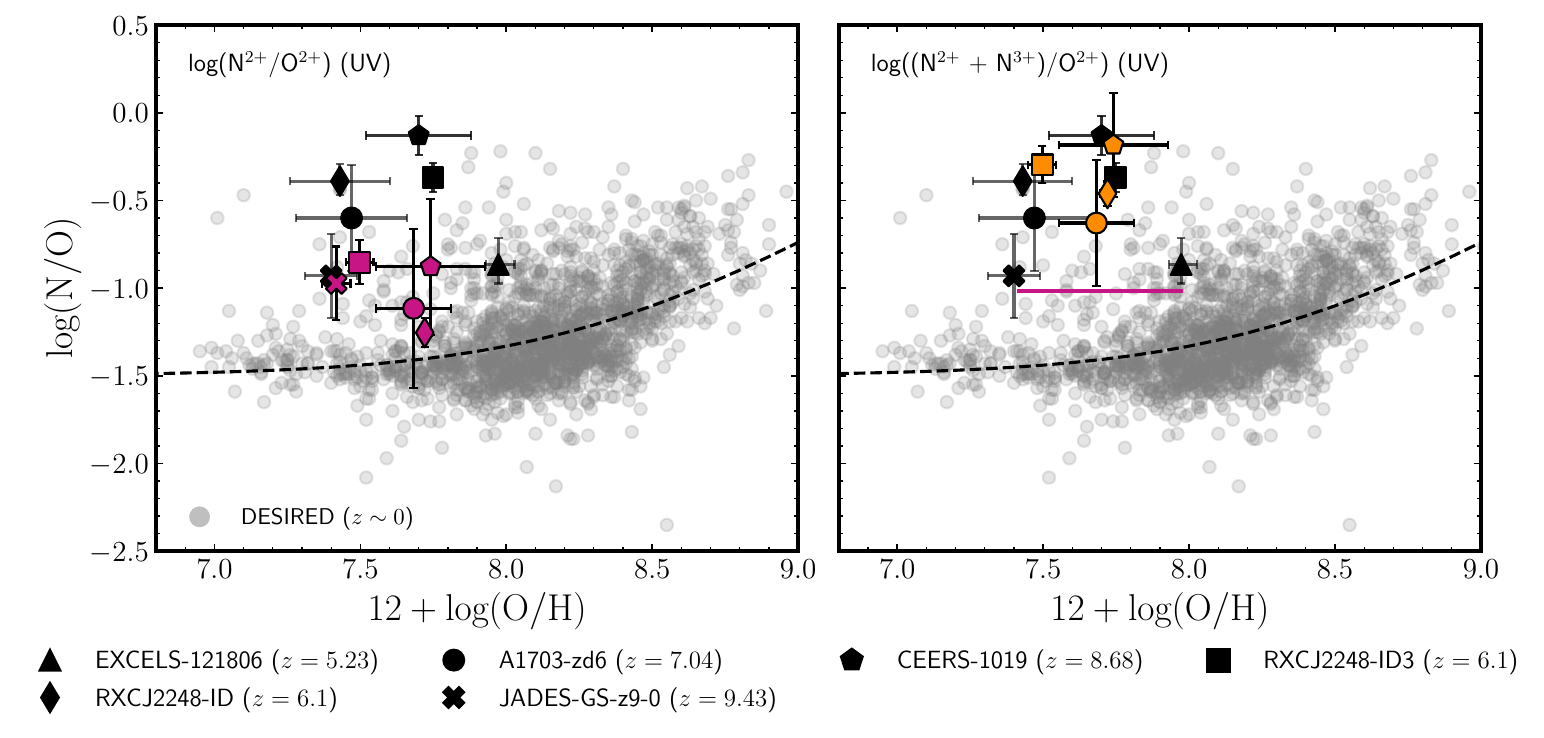}
    \caption{The N/O–O/H relation for galaxies at $z > 5$ is shown using different symbols. Black symbols correspond to the original abundance ratios reported by the reference authors, which exhibit signatures of nitrogen enrichment at $12 + \log(\mathrm{O/H}) < 8.0$ \citep{topping24a, topping25a, arellanocordova25, marqueschaves:24, Berg25b}. The N/O ratios reported in \citet{Berg25b} correspond to values obtained without applying an ICF correction (see Sec.~\ref{sec:disc}). For reference, local star-forming regions from the DESIRED project \citep{MendezDelgado:23b, esteban25}, with abundances determined via the direct method, are shown as gray circles. The O/H and N/O, (N$^{2+}$/O$^{2+}$) and  (N$^{2+}$ + N$^{3+}$)/O$^{2+}$), re-derived using our self-consistent procedure described in Sec.~\ref{sec:physica_conditions} are presented as pink (left panel) and orange (right panel) symbols, respectively. For reference, the right panel shows the mean value of $\log(\mathrm{N^{2+}/O^{2+}}) = \log(\mathrm{N/O}) = 1.01 \pm 0.17$, indicated by a dashed line. The revised N/O estimates using N$^{2+}$/O$^{2+}$ for high-$z$ galaxies are consistent with the intrinsic dispersion observed in the local sample, whereas the inclusion of N$^{3+}$ increases the N/O ratio by 0.64 dex relative to the N$^{2+}$/O$^{2+}$ ratio. The dashed line represents the relation of \citet{nicholls17} from stellar abundances from the Milky Way.  The results of N/O derived using N$^+$/O$^{+}$ and (N$^+$ + N$^{2+}$)/O$^{2+}$) are discussed in Sec.~\ref{sec:disc}.}
    \label{fig:NO}
\end{figure*}

\begin{figure}
    \centering
    \includegraphics[width=0.90\linewidth]{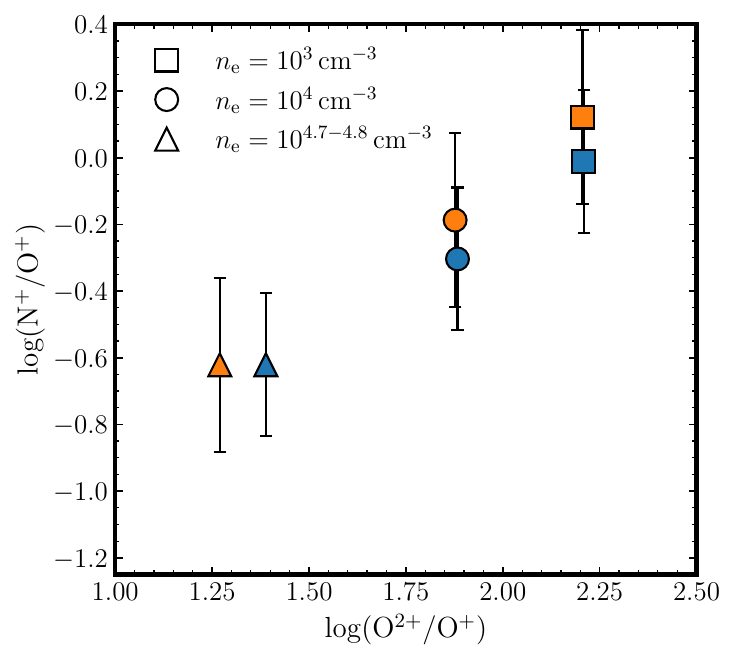}
    \caption{Results for the N$^{+}$/O$^{+}$ ratio as a function of O$^2+$/O$+$, derived using densities of 10$^{3}$ cm$^{-3}$ (squares), 10$^{4}$ cm$^{-3}$ (circles), and $10^{4.7-4.8}$ cm$^{-3}$ (triangles) for RXCJ2248-ID3-T24 (blue) and RXCJ2248-ID3-B25 (orange). The higher densities in this analysis ({$10^{4.7-4.8}$ cm$^{-3}$}, triangles) correspond to the \siliii] values reported in the original studies of \citet{topping24a} and \citet{Berg25b}. The correlation between N$^{+}$/O$^{+}$ and O$^{2+}$/O$^{+}$ arises from the low critical density of [\oii]$\lambda\lambda$3726,29, which leads to an overestimation of O$^{+}$ as the density increases.  Note that this illustrative result shows the impact of the variations of N$^{+}$/O$^{+}$ when assuming three different values of the electron density. The inferred N$^{+}$/O$^{+}$ ratios for RXCJ2248-ID-T24 and RXCJ2248-ID3-B25 correspond to the use of \Ne$=10^{4}$ cm$^{-3}$ for low ionization ions as our preferred values. The results in Table~\ref{tab:oiii_properties} shows the final N$^{+}$/O$^{+}$ ratio inferred as indicated in Sec.~\ref{sec:physica_conditions} for each galaxy.}
    \label{fig:NO_optico}
\end{figure}

In Fig.~\ref{fig:NO}, we present the N/O–O/H relation for the selected high-$z$ galaxies for log(N$^{2+}$/O$^{2+}$) (left) and log((N$^{2+}$ + N$^{3+}$)/O$^{2+}$) (right), together with the local DESIRED sample. The results from the original studies are also included and shown as black symbols. For most objects, the N/O abundances were derived primarily from UV lines, whereas for EXCELS-121806 and RXCJ2248-ID-T24, and  RXCJ2248-ID3-B25 an optical determination of N/O was also possible \citep[][]{arellanocordova25, Berg25b}.
For EXCELS-121806, RXCJ2248-ID-T24 and RXCJ2248-ID-B25, we obtain slightly high values of $\log(\mathrm{N/O}) = -0.96 \pm 0.25$, $-0.30 \pm 0.21$ and $-0.18 \pm 0.26$, respectively, from rest-frame optical N$^+$ and O$^+$ lines. 
In Fig.~\ref{fig:NO_optico}, we also show the $\log(\mathrm{N/O})$ ratio as a function of the ionization degree, O$^{2+}$/O$^{+}$, for RXCJ2248-ID3-T24 and RXCJ2248-ID3-B25, where N$^{+}$ and O$^{+}$ are determined using \Ne=$10^3-10^4$ cm$^{-3}$.
For the \Ne\ = $10^{4.7-4.8}$ cm$^{-3}$ implied by the \ion{Si}{iii}], as reported in \citet{topping24a} and \citet{Berg25b}, $\log(\mathrm{N/O})$ decreases because O$^{+}$ becomes overestimated, a consequence of the low critical density of [\oii]~$\lambda\lambda$3726,29. Therefore, O$^{+}$ calculated under these conditions may critically affect the derived total N/O abundance (or the O$^{2+}$/O$^{+}$ ratio). In contrast, the impact on the metallicity depends on whether O$^{+}$ is also a dominant ion on the total abundance of O/H \citep[e.g.,][]{marqueschaves:24, arellanocordova25b}. Note that the inferred values of log(N$^{+}$/O$^{+}$), assuming \Ne$=10^{4}$ cm$^{-3}$ in Fig.~\ref{fig:NO_optico}, correspond to the results listed in Table~\ref{tab:oiii_properties} (see also Sec.~\ref{sec:physica_conditions}) for RXCJ2248-ID3-T24 and RXCJ2248-ID3-B25 as our preferred values. This electron density is assumed in the computation of low-ionization ions.

On the other hand, when the N/O ratio is derived from the \niii]/[\oiii] lines (N$^{2+}$/O$^{2+} \sim$ N/O), the results are more consistent with the DESIRED expectations given the metallicities of these galaxies, falling in the range $-1.25 < \log(\mathrm{N/O}) < -0.97$ (see left panel of Fig.~\ref{fig:NO}). The apparent overestimation of N/O inferred from optical lines may result from adopting electron densities determined with classical diagnostics such as [\sii]~$\lambda 6717/\lambda6731$. Indeed, \citet{MendezDelgado:23b} present compelling evidence that this diagnostic systematically underestimates densities in the presence of density variations. If such variations are evident in the local universe, it is reasonable to expect them at high redshift as well. Adopting higher electron densities in the determination of optical N/O abundances leads to correspondingly lower N/O values.
To achieve consistency between N$^{+}$/O$^{+} \approx$ N$^{2+}$/O$^{2+}$ in RXCJ2248-ID, the gas density in the low-ionization zone must also be of the same order of magnitude as that inferred from \Ne([\oiii]). This would imply that the density in the low-ionization region can be much higher than what can be deduced from classical indicators such as [\sii]~$\lambda 6717/\lambda 6731$. Such a scenario is in fact expected: if the density is significantly higher than $10^{3}$ cm$^{-3}$, collisional de-excitation dominates in both atomic levels that give rise to [\sii]~$\lambda \lambda 6717,6731$, effectively suppressing radiative de-excitation. It would be naive to expect evidence of high density from a diagnostic based on atomic levels that cannot produce a significant number of radiative transitions in dense regions. Moreover, one can hardly expect photoionization conditions where N$^{+}$/N$^{2+}$ is extremely different from O$^{+}$/O$^{2+}$, which reinforces the hypothesis of substantial density variations in the low-ionization zone.

The potential bias in O$^{2+}$/O$^{+}$, arising from the propagation of uncertainties in the determination of the electron density due to internal variations, represents a serious problem for all chemical abundances that rely on ICFs based on this ratio. In particular, the abundances of elements such as Ar require the use of an ICF, and recent studies have reported anomalies in the distribution of Ar/O versus O/H at high redshift \citep[e.g.,][]{rogers24,Bhattacharya:2025} compared to the trends observed in the local universe \citep[e.g.,][]{arellanocordova25b, esteban25}. Analogous to the case of N/O versus O/H, these results have led to proposals of alternative enrichment pathways for high-$z$ galaxies, such as rapid but intermittent star formation \citep{Bhattacharya:2025}. Although a detailed analysis of Ar/O versus O/H abundances is beyond the scope of this paper, we emphasize that the proper determination of density and temperature in the high-density regime observed at high redshift has a critical impact on the determination of the abundances of all elements.

Although the N/O abundances inferred from \niii] and \oiii] UV lines can change significantly (whereas those derived from optical diagnostics may decrease if higher densities are adopted), Fig.~\ref{fig:NO} shows that there may also be an important increase in O/H, which traces the global metallicity. For RXCJ2248-ID (diamond) and A1703-zd6 (circle), the metallicities increase by $\sim$0.29 dex when using our self-consistent method. With this increase, the N/O ratios derived from N$^{2+}$/O$^{2+}$ become fully consistent with those expected for local star-forming regions in the DESIRED sample \citep[see also][]{berg18, hayes25, martinez25}. 

Up to this point, we have discussed the N/O values inferred under the assumptions N$^{2+}$/O$^{2+} \approx$ N/O and N$^{+}$/O$^{+} \approx$ N/O. As highlighted in studies focused on the derivation of ICFs \citep[e.g.,][]{amayo21}, these aforementioned ionic abundance ratios exhibit intrinsically large dispersions and thus provide, at best, crude approximations to the total N/O abundance \citep{martinez25}. Obtaining a robust ICF for nitrogen in high-redshift galaxies requires photoionization models tailored to their physical conditions, an effort that, as noted above, is still under active development and lacks consensus in the current literature. Ideally, the most reliable N/O determination involves summing the ionic contributions N$^{+}$, N$^{2+}$, and N$^{3+}$ for nitrogen, and O$^{+}$, O$^{2+}$, and O$^{3+}$ for oxygen.

Photoionization models from the Mexican Million Models database (3MdB; \citealt{Morisset:2015}) predict that the contribution of O$^{3+}$ is negligible for the ionization degrees implied by the observed O$^{2+}$/O$^{+}$ ratios in the galaxies analyzed here. Under these circumstances, the approximation  $\frac{N^{+}+N^{2+}+N^{3+}}{O^{+}+O^{2+}} \approx$ N/O is reasonable. One must keep in mind, however, that \citet{RickardsVaught25} has shown that classical photoionization models can underestimate O$^{3+}$ abundances by up to a factor of two in extremely metal-poor systems such as I Zw 18 when compared with observations.

The N/O abundances obtained when including the N$^{3+}$ abundances inferred from the UV \niv] lines in RXCJ2248-ID-T24, A1703-zd6, and CEERS-1019 are exceptionally high compared to the values derived from N$^{2+}$/O$^{2+}$ and/or N$^{+}$/O$^{+}$ (see Fig.~\ref{fig:NO}). It is crucial to emphasize that, in these cases, the total nitrogen abundance is overwhelmingly dominated by N$^{3+}$/H$^{+}$; that is, the extreme N/O ratios arise almost entirely from the inferred N$^{3+}$ abundance, with the other ions contributing negligibly.

This interesting behavior requires a discussion of how N$^{3+}$/H$^{+}$ is derived from the \niv] lines before incorporating it into the total N/O estimate. Beyond the challenges associated with ICFs, such as the possibility that the N$^{3+}$-emitting region does not fully overlap with the O$^{2+}$ zone, which would lead to an overestimate of N/O, the key issue lies in the conversion of the observed \niv] intensities into ionic abundances. This conversion is strongly sensitive to the electron temperature due to the exponential dependence of the emissivity on the excitation energy (8.33 eV) of the $^3P$-N$^{3+}$ levels, which is more than 50 per cent higher than that of the [\oiii]~$\lambda 4363$ transition, a line well known for its temperature sensitivity.

The ionic N$^{3+}$/H$^{+}$ abundances listed in Table~\ref{tab:oiii_properties} and those from the reference studies assume that the electron temperature of the O$^{2+}$ zone, \Te([\oiii]), is also representative of N$^{3+}$, even though the latter belongs to a higher ionization regime. This assumption has not been observationally verified with diagnostics tracing higher-ionization temperatures, such as \Te([\ion{Ar}{iv}]~$\lambda\lambda 7170^+/\lambda\lambda 4720^+$) or \Te([\ion{Ar}{v}]~$\lambda 4625/\lambda 7005$). While classical photoionization models predict relatively small deviations (typically $\sim$5 per cent) between \Te(O$^{2+}$) and \Te(N$^{3+}$) in chemically homogeneous nebulae, this may change dramatically if a substantial fraction of the N$^{3+}$ emission arises from a physically distinct volume ---for example, a nitrogen-rich, highly ionized outflow or ejecta. The choice of electron temperature is therefore critical: a mere $\sim$10 per cent increase in \Te(N$^{3+}$) relative to \Te([\ion{O}{iii}]) would reduce the inferred N$^{3+}$/O$^{2+}$ abundance by roughly 50 per cent. Conversely, if \Te(N$^{3+}$) were slightly lower than \Te([\ion{O}{iii}]), the inferred N$^{3+}$/O$^{2+}$ ratio would increase accordingly. This illustrates that even modest temperature stratification can introduce large systematic biases in the derived abundance ratio.

In summary, the interpretation of the strong \niv]~$\lambda\lambda1483,1486$ lines is far from straightforward. If one assumes the validity of the empirically untested hypothesis \Te([\oiii]) $\approx$ \Te(N$^{3+}$), then the observations imply very high nitrogen abundances dominated by this particular ionization state, and one must explain why N/O is so elevated at a fixed O/H, as shown in the right panel of Fig.~\ref{fig:NO}. This is the research line followed in several recent studies referenced here. Alternatively, the physical conditions in the ionized gas of high-$z$ galaxies may depart from local expectations, with \Te(N$^{3+}$) being somewhat higher than \Te([\ion{O}{iii}]), in which case the total N/O abundance would not be as anomalous as previously thought. This latter scenario is plausible, especially considering that our analysis finds significant variations in electron density across different ionization zones. Since density and temperature are coupled, strong variations in the former could induce variations in the latter. A combination of mildly enhanced N/O relative to local values and temperature stratification also remains as a possible interpretation.

It is important to emphasize that the local DESIRED local sample spans a wide parameter space in N/O and O/H, implying an intrinsic dispersion when compared to results derived from strong-line methods, which tend to artificially reduce such dispersion \citep[][see their fig.~1]{valeasari16}. Although strong-line calibrations are anchored to samples of objects with \Te\ determined via the direct method \citep[e.g.,][]{dopita+16, sanders23b, scholte25}, these calibrations can introduce significant biases in the determination of N/O and metallicity.

In addition, we want to be emphatic: at present, photoionization and shock models for high-$z$ galaxies observed with JWST are still under active development. Although analogs can be found with conditions observed in local systems \citep[e.g.,][]{mingozzi22, flury25}, several key physical parameters remain under investigation, including the underlying stellar population, the density and temperature structure, and the dust composition, distribution and geometry. This is particularly relevant because determining N/O abundances requires assumptions about the coexistence of different ionic states of both elements—that is, the adopted ionization correction factors (ICF). More importantly, it relies on the ionic abundances being accurately determined, which in turn requires a reliable determination of the electron density and temperature for each ionization zone. Indeed, even if the correct ICF is adopted to derive N/O, any systematic error in one of the ionic abundances involved would inevitably propagate into the final result.

\subsection{Current Limitations and Directions for Future Work}

The methodology proposed in this work is intended as a turning point in the critical analysis of the physical and chemical conditions of ionized gas in high-$z$ galaxies, rather than as a simple ``recipe'' for determining the chemical composition of the universe at high-$z$. A self-consistent approach, such as the one presented here, is certainly desirable, as it reduces the number of assumptions required to infer the chemical composition of the gas. The combined use of \oiii]~\W1664$^{+}$ and [\oiii]~\W\W 4363, 5007 allows us to avoid assuming the coexistence of different ions to determine the abundance of O$^{2+}$ --the most relevant ionic state of oxygen in high-$z$ galaxies-- and has revealed important differences with ``classical'' methodologies employed in previous reference studies. However, the inconsistency among the different density diagnostics present in the ionized gas indicates internal complexities in its distribution. Indeed, in the case of an inhomogeneous density structure, each diagnostic will be biased according to the critical density of the atomic levels involved \citep{Rubin:89,  MendezDelgado:23b}. Thus, each diagnostic carries its own bias governed by its sensitivity range.

The existence of density inhomogeneities in ionized gas has been firmly established since the pioneering spectroscopic studies of the local universe \citep{Seaton:57}. These variations are not small but can span several orders of magnitude, giving rise to the well-known discrepancies between densities derived from \hi\ line fluxes (commonly referred to as \textit{rms} densities) and those inferred from forbidden-line fluxes, which in turn led to the physical concept of the filling factor. Although recently some authors have argued in favor of a homogeneous density structure in the local universe \citep[e.g.,][]{Chen:23, Harikane:25}, the observational evidence strongly suggests that this scenario is unlikely \citep[e.g.,][]{MendezDelgado:24a, RickardsVaught24, Yarovova:25}. The results presented in this work seem to indicate that significant density variations are also present in the high-$z$ universe, but likely at systematically higher density regimes, capable of biasing diagnostics such as [\oiii]~$\lambda 4363/5007$. By contrast, in the local universe such variations may still span several orders of magnitude but typically occur at low densities.

Nevertheless, even these internal density variations in a lower density regime are sufficient to affect some temperature diagnostics. For example, \citet{MendezDelgado:23b} show that line diagnostics such as [\oii]$\lambda \lambda 7330^+/3727^+$ yield systematically higher temperatures than other diagnostics such as [\nii]$\lambda 5755/6584$ when densities are estimated from classical ratios like [\sii]$\lambda 6717/6731$, closely analogous to the case presented here for [\oiii]. Although these analyses have been carried out in the local universe, the fundamental principles of ionized-gas analysis are identical to those applied to high-$z$ galaxies, where auroral [\oii]~\W\W 7320, 7330 lines have also been detected \citep[e.g.,][]{sanders23a, cataldi25, sanders:25}. It is therefore very likely that the use of these lines to determine \Te([\oii]) at high redshift will introduce strong systematic biases related to density.

Given that, in the presence of density inhomogeneities, all diagnostics may carry some degree of systematic bias depending on their sensitivity ranges, it follows that our self-consistent method may also be subject to bias. Determining whether such a bias exists --and, if so, whether it shifts the inferred densities toward higher or lower values (which has a significant impact on the derived temperatures)-- requires a multi-diagnostic analysis in a similar way than for objects in the local universe as done by \citet{MendezDelgado:23b}. This, in turn, demands a larger number of JWST observations with simultaneous detections of \oiii] $\lambda \lambda1664^{+}$ and [\oiii]$\lambda \lambda 4363, 5007$ in addition to other diagnostics such as [\sii]~$\lambda 6717/6731$, \ciii]~$\lambda 1909/1907$, and others, implying the need to re-observe multiple high-$z$ galaxies.

An additional issue to consider is the possibility of genuine electron temperature variations. Indeed, an inhomogeneous density structure in the high-density regime can bias the interpretation of the [\oiii]$\lambda4363/5007$ ratio toward higher temperatures due to collisional de-excitation of the $^1D_2$ level that produces [\oiii]~$\lambda 5007$ \citep[e.g.,][]{marconi24, hayes25}. However, an inhomogeneous temperature structure can also bias this ratio because the $^1S_0$ level has a stronger dependence on \Te\ owing to its higher excitation energy, as originally predicted by \citet{peimbert67}. Some authors have presented evidence for temperature inhomogeneities in star-forming regions, which bias CEL-based chemical abundances toward values systematically lower than the true ones, in close analogy to the situation discussed here \citep[e.g.,][]{garcia-rojas07, MendezDelgado:23a}.

From a thermodynamic standpoint, it seems implausible to invoke strongly inhomogeneous density structures while simultaneously assuming a homogeneous temperature structure, given the intrinsic coupling between these two quantities. A more comprehensive study must therefore also assess the impact of genuine temperature variations on abundance determinations at high-$z$, in conjunction with density effects \citep[e.g.,][]{MendezDelgado:23a, marconi24, hayes25, MendezDelgado:25}. Furthermore, processes such as fluorescence through permitted transitions connecting high-lying atomic levels to the ground state could also affect the derived temperatures, although such effects are expected to be more prominent in metal-rich galaxies and therefore with lower global mean electron temperature (Morisset et al. in prep.). 

As previously discussed, density-related biases affect not only the total N/O and O/H abundances but also ionic ratios such as O$^{2+}$/O$^{+}$, thereby propagating errors into all chemical abundances that require an ICF to account for unseen ionic fractions. A clear example is Ar/O, for which “anomalous” cases have also been reported at high-$z$. It is therefore essential to carefully examine the physical assumptions underlying abundance determinations before interpreting chemical anomalies that may, in fact, arise from systematic errors in the estimation of the physical conditions.

Finally, it is important to emphasize the impact that the choice of atomic data can have on line diagnostics. Several studies have shown that uncertainties in radiative and collisional coefficients for optical lines become particularly significant when densities exceed $10^4$ cm$^{-3}$, leading to systematic biases in O and N/O abundances of up to 0.8 dex \citep{JuandeDios:17, Morisset:20, Mendoza:23}. At high redshift, this problem is compounded: on the one hand, we are dealing with densities well above this threshold, and on the other, there is a clear lack of systematic studies evaluating the reliability of atomic parameters for the higher energy levels that give rise to UV lines. Our intention here is to highlight these unresolved, fundamental issues before invoking the existence of exotic conditions in the high-$z$ universe based on chemical abundance determinations that may not yet be sufficiently robust.

\section{Summary and Conclusion}
\label{sec:conclusion}

We present a self-consistent methodology for determining the physical conditions of high-$z$ galaxies, applying a robust direct-method approach that simultaneously combines the \Te-\Ne\ diagnostics [\oiii]$\lambda 4363/\lambda 5007$ and \oiii]$\lambda 1666$/\oiii]$\lambda 4363$ to derive \Ne([\oiii]) and \Te([\oiii]). With these determinations, we infer the metallicities and nitrogen abundances of four high-$z$ galaxies, comparing our results with a large sample of local  SFGs and extragalactic \ion{H}{ii} regions from the DESIRED project. Our main conclusions are:

\begin{itemize}

\item [1.] Using our self-consistent [\ion{O}{iii}]-based method, we recover electron densities that are, in most of the cases, systematically higher than those assumed in the literature. In particular, we find \Ne([\oiii]) in the range $0.3$–$3.0 \times 10^{5}$ cm$^{-3}$ with high electron temperatures (\Te $\sim 20,000$ K).

\item [2.] Accounting for high densities reduces the derived electron temperatures and viceversa, which in turn raise the inferred metallicities in some galaxies. This mitigates part of the apparent offset between $z > 5$ galaxies and local systems in the MZR, suggesting a more continuous chemical evolution across cosmic time. 

\item[3.] Reports of anomalously high N/O ratios at early epochs in the analyzed sample rely on the high N$^{3+}$/H$^{+}$ abundances. In most cases, when robust \Te\ and \Ne\ estimates are employed, the discrepancies in the N/O–O/H relation decrease as O/H increases. Furthermore, we argue that a temperature stratification in which \Te(N$^{3+}$) is somewhat higher than \Te([\oiii]) could substantially reduce the elevated N/O abundances inferred primarily from the \niv]~$1483,86$ lines, which are extremely sensitive to \Te. 

Our results emphasize that density effects must be explicitly accounted for in direct-method abundance studies of high-redshift galaxies. Incorporating these corrections is crucial for reliable metallicity and N/O determinations. This is also relevant for elements whose abundances rely on ICFs such as Ar. While our approach provides a robust framework, it does not explicitly account for intrinsic temperature fluctuations within the nebulae. In practice, strong density variations without associated changes in \Te\ seems unlikely, and such coupled fluctuations may still bias abundance determinations, particularly at the precision now achievable with JWST. 

Finally, the reliability of abundance determinations ultimately depends on the accuracy of atomic data at high densities. Further improvements in collision strengths and transition probabilities for key ions are critical to refine metallicity and N/O estimates in galaxies at early cosmic times. Local and lower redshifts star-forming galaxies \citep[e.g., the Sunburst Arc lensed galaxy that also shows extremely high N/O,][]{pascale23}, with available UV and optical spectra will benefit from the methodology presented in this study, particularly in cases where high-density clumps might be presented \citep[e.g.][]{mingozzi22, jung25,martinez25}. However, we stress the carefully treatment of other source of uncertainties local systems with no simultaneous detections of [\ion{O}{iii}] and \ion{O}{iii}] due to aperture affects, flux calibration, and dust extinction.
Extending this analysis to larger and more diverse samples is necessary to establish whether the trends identified here are representative of the broader high-$z$ galaxy population. Forthcoming JWST programs combining UV and optical diagnostics will provide the crucial data needed to generalize and test the robustness of this methodology.  

\end{itemize}

\section*{Acknowledgments}
The authors thank the anonymous referee for their thoughtful feedback that improved this paper.
KZAC, FC, TMS, and DS support from a UKRI Frontier Research Guarantee Grant (PI Cullen; grant reference: EP/X021025/1). 

JEMD, LC, CM and AP thanks the support by SECIHTI CBF-2025-I-2048 project ``Resolviendo la Física Interna de las Galaxias: De las Escalas Locales a la Estructura Global con el SDSS-V Local Volume Mapper'' (PI: Méndez-Delgado). 

KZAC, JEM-D, LC, CE, JGR, FFRO, CM and AP thank the support by UNAM/DGAPA/PAPIIT/IA103326 project ``DESIRED (DEep Spectra of Ionized Regions Database): de las emisiones más sutiles a la física fundamental del universo’’ (PI: Méndez-Delgado).

 Based on observations with the NASA/ESA/CSA James Webb Space Telescope obtained at the Space Telescope
Science Institute, which is operated by the Association of Universities for Research in Astronomy, Incorporated, under NASA356
contract NAS5-
03127. 

CE and JGR acknowledge financial support from the Agencia Estatal de Investigaci\'on of the Ministerio de Ciencia, Innovaci\'on y Universidades under grant ``The internal structure of ionised nebulae and its effects in the determination of the chemical composition of the interstellar medium and the Universe'' with reference PID2023-151648NB-I00 (DOI:10.13039/5011000110339). JGR also acknowledges financial support from AEI-MCIU and from the European Regional Development Fund (ERDF) under grant ``Planetary nebulae as the key to understanding binary stellar evolution'' with reference PID-2022136653NA-I00 (DOI:10.13039/501100011033).

KK gratefully acknowledges funding from the Deutsche Forschungsgemeinschaft (DFG, German Research Foundation) in the form of an Emmy Noether Research Group (grant number KR4598/2-1, PI Kreckel) and the European Research Council’s starting grant ERC StG-101077573 (“ISM-METALS").

This work was initiated and discussed as part at the Aspen Center for Physics, which is supported by a grant from the Simons Foundation (1161654, Troyer).

\section*{Data Availability}
All data used in this study is available in the original studies where the fluxes were compiled. 


%
    Software: \texttt{jupyter} \citep{kluyver16}, \texttt{astropy} \citep{astropy:2018, astropy:2022}, {\tt PyNeb} \citep{luridiana15}, {\tt matplotlib} \citep{matplotlib},
{\tt numpy} \citep{numpy}, {\tt scipy} \citep{scipy}.



\bibliographystyle{mnras}
\bibliography{refs} 



\appendix
\section{Atomic data and critical densities}
\label{sec:atomic}

\begin{figure}
    \centering
    \includegraphics[width=\linewidth]{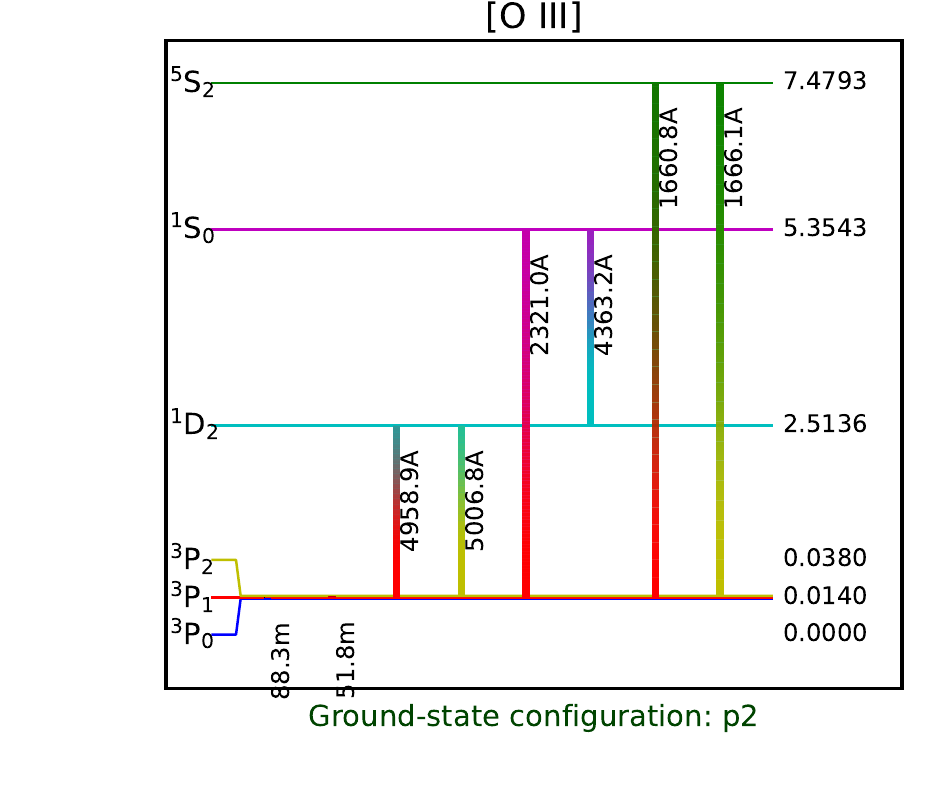}
    \caption{Grotrian diagram \citep{Grotrian:28} illustrating the forbidden transitions of O$^{2+}$, generated with {\tt PyNeb} \citep{luridiana15}.}
    \label{fig:O3_Grotrian}
\end{figure}

In Fig.~\ref{fig:O3_Grotrian}, we present the Grotian diagram  \citep{Grotrian:28}, illustrating the \oiii\ transitions employed in this study to derive \Te\ and \Ne\ in a self-consistent manner. The adopted atomic data are listed in Table~\ref{tab:atomic_data}, selected following previous analyses of the physical conditions and chemical properties of star-forming regions \citep[e.g.,][]{MendezDelgado:23b, esteban25}. For reference, Table~\ref{tab:O3_levels} provides the critical densities of the relevant \oiii\ transitions, further emphasizing the robustness of the methodology adopted in this work.

\begin{table}
\centering
\footnotesize
\caption{Atomic data used in this work}
\label{tab:atomic_data}
\begin{tabular}{l c c}
\toprule
Ion & Transition probabilities ($A_{ij}$) & Collision strengths ($\Upsilon_{ij}$) \\
\midrule

O$^{+}$   & \citet{fft04}        & \citet{Kisielius:2009} \\
O$^{2+}$  & \citet{storey00}        & \citet{aggarwal99}  \\    
          & \citet{wiese96}        &      \\
N$^{+}$   & \citet{fft04}        & \citet{tayal11}        \\
N$^{2+}$   & \citet{galavis98}     & \citet{blum92} \\
N$^{3+}$   & \citet{wiese96}        & \citet{ramsbottom94} \\

\bottomrule
\end{tabular}
\end{table}


\begin{table}
\centering
\caption{Reference information on the O$^{2+}$ state properties and transitions. Critical densities $n_{\rm crit}$ assume a temperature of $10^4$ K using the atomic data for O ions listed in Table~\ref{tab:atomic_data} }
\label{tab:O3_levels}
\begin{tabular}{l c c c c}
\toprule
Wavelength (\AA) & Lower level & Upper level & $\log n_{\rm crit}$ (cm$^{-3}$) \\
\midrule
 1660.809 & $\rm ^3P_1$ & $\rm ^5S_2$ & 10.615 \\
1666.150 & $\rm ^3P_2$ & $\rm ^5S_2$ & 10.615 \\
 4363.209 & $\rm ^1D_2$ & $\rm ^1S_0$ & 7.336 \\
4958.910 & $\rm ^3P_1$ & $\rm ^1D_2$ & 5.820 \\
 5006.842 & $\rm ^3P_2$ & $\rm ^1D_2$ & 5.820 \\
\bottomrule
\end{tabular}
\end{table}


\section{Monte Carlo Simulations}
\label{sec:MC_simulations}
\begin{figure}
    \centering
    \includegraphics[width=\linewidth]{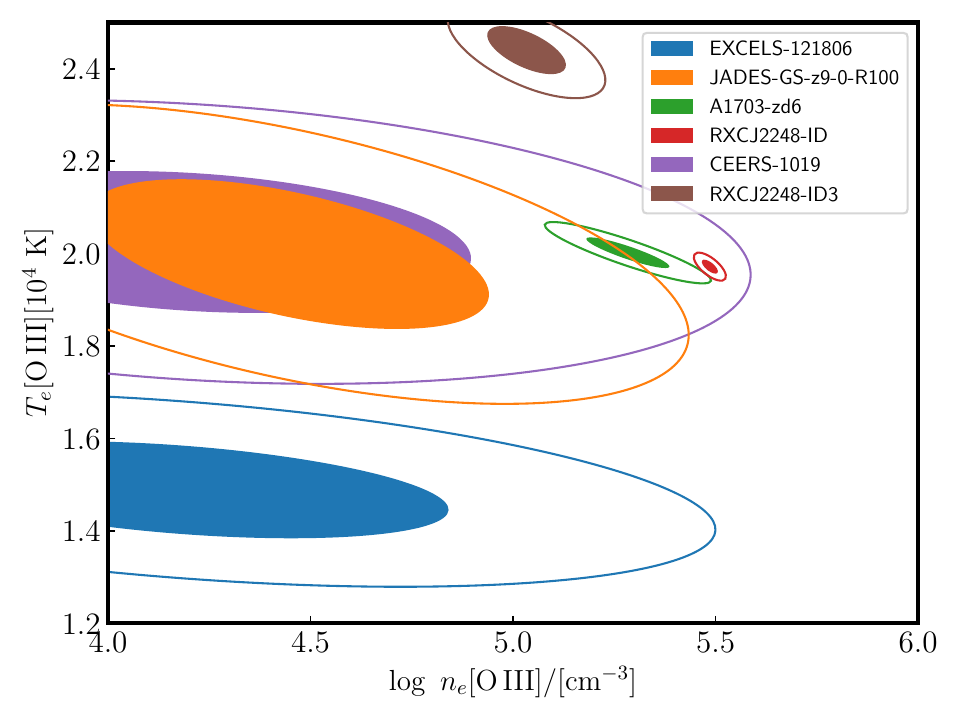}
    \caption{The 1- and 2-$\sigma$ confidence intervals (filled and open ellipses, respectively) on \Te\ and \Ne\ from forward modeling of the observed line ratios by MCMC sampling of the posterior on temperature and density. The diagnostic power of well-detected lines is evident for RXCJ228-ID-T24, RXCJ228-ID3-B25, and A1703-zd6.}
    \label{fig:MC_simulations}
\end{figure}

To forward-model the observed line ratios, we perform a nonlinear least squares fit of the emissivity ratios predicted by {\tt PyNeb} to the observed flux ratios, with \Te\ and \Ne\ as free parameters. We then seed the {\tt emcee} MCMC sampler with these values and sample the posterior on \Te\ and \Ne\ with $10^4$ steps for each of 5 walkers, with an additional $2\times10^3$ steps for burn-in to forget the initial seed. For the MCMC sampler, we assume a log likelihood of $\chi^2$, i.e., the variance-weighted summed squared difference between the predicted and measured flux ratios. We further impose on \Ne\ a heuristic prior of
\begin{equation}
    p(n{\rm_e}) = 1-\exp\left(-\frac{n{\rm_e}}{5852\rm~cm^{-3}}\right),
\end{equation}
such that $p(n{\rm_e})<0.5$ when $n{\rm_e}<n_{\rm crit}$ for the [\ion{S}{ii}] and [\oii] doublets ($n_{crit}\sim3\,000\rm~cm^{-3}$ and $\sim4\,000\rm~cm^{-3}$, respectively). In the case of the objects considered here, these doublets are saturated towards the high density limit of their diagnostic range, disfavoring the low density regime to which these lines are sensitive \citep[cf.][]{MendezDelgado:23b}. We note that this heuristic prior only affects the posterior sampling of CEERS 1019, JADES-GS-z9, and EXCELS-121806 as these three sources have [\oiii] 5007/4363 ratios in the low-density limit (see also Fig.~\ref{fig:plasmaplots}). Otherwise, a uniform prior gives results similar to the heuristic prior.


\bsp	
\label{lastpage}
\end{document}